\documentstyle[12pt,epsf,axodraw]{article}

\begin{document}

\baselineskip 14pt

\newcommand{\sheptitle}
{Dipole moments of the Electron, Neutrino and Neutron
in the MSSM without R-parity Symmetry} 

\newcommand{\shepauthor}
{S. A. Abel$^a$ and A. Dedes$^b$ and H. K. Dreiner$^b$}

\newcommand{\shepaddress}
{$^a$Service de Physique Th\'eorique, CEA-SACLAY, 
Gif-sur-Yvette, 91191 France\\
 $^b$Rutherford Appleton Laboratory, Chilton, Didcot, OX11 0QX, UK}

\newcommand{\shepabstract}
{We show that in the MSSM without R-parity symmetry there are {\em no}
new contributions to electron and neutron electric dipole moments
(EDMs) at 1-loop induced by the R-parity violating Yukawa
couplings. Non-zero EDMs for the electron and neutron first arise at
the 2-loop level. As an example we estimate the contribution of a
two-loop graph which induces electron EDMs.  On the other hand, we
show that the (Majorana) neutrino electric and magnetic transition
moments are non-zero even at the 1-loop level. Constraints on the
R-parity violating couplings are derived from the existing bounds on
the neutrino dipole moments. }

\begin{titlepage}
\begin{flushright}
hep-ph/9912429\\
RAL-TR-1999-87\\
\end{flushright}
\vspace{0.5in}
\begin{center}
{\large{\bf \sheptitle}}
\bigskip \\ \shepauthor \\ \mbox{} \\ {\it \shepaddress} \\ 
\vspace{0.5in}
{\bf Abstract} \bigskip \end{center} \setcounter{page}{0}
\shepabstract
\vspace{0.5in}
\begin{flushleft}
\today
\end{flushleft}
\end{titlepage}

\def\sspace{\baselineskip = .16in}
\def\dspace{\baselineskip = .30in}
\def\beq{\begin{equation}}
\def\eeq{\end{equation}}
\def\bea{\begin{eqnarray}}
\def\eea{\end{eqnarray}}
\def\bq{\begin{quote}}
\def\eq{\end{quote}}
\def\ra{\rightarrow}
\def\lra{\leftrightarrow}
\def\ups{\upsilon}
\def\bq{\begin{quote}}
\def\eq{\end{quote}}
\def\ra{\rightarrow}
\def\un{\underline}
\def\ov{\overline}

\newcommand{\plb}[3]{{{\it Phys.~Lett.}~{\bf B#1} (#3) #2}}
\newcommand{\npb}[3]{{{\it Nucl.~Phys.}~{\bf B#1} (#3) #2}}
\newcommand{\prd}[3]{{{\it Phys.~Rev.}~{\bf D#1} (#3) #2}}
\newcommand{\prl}[3]{{{\it Phys.~Rev.~Lett.}~{\bf #1} (#3) #2}}
\newcommand{\ptp}[3]{{{\it Prog.~Theor.~Phys.}~{\bf #1} (#3) #2}}
\newcommand{\leqsim}{\,\raisebox{-0.6ex}{$\buildrel < \over \sim$}\,}
\newcommand{\geqsim}{\,\raisebox{-0.6ex}{$\buildrel > \over \sim$}\,}
\newcommand{\be}{\begin{equation}}
\newcommand{\ee}{\end{equation}}
\newcommand{\ba}{\begin{eqnarray}}
\newcommand{\ea}{\end{eqnarray}}
\newcommand{\nn}{\nonumber}
\newcommand{\dd}{\mbox{{\rm{d}}}}
\newcommand{\ie}{\mbox{{\em i.e.~}}}
\newcommand{\eg}{\mbox{{\em e.g.~}}}
\newcommand{\mpl}{\mbox{$M_{pl}$}}
\def\gev{\,{\rm GeV}}

\section{Introduction}

The electric dipole moment (EDM) $d_f$, and magnetic dipole moment
(MDM) $\mu_f$, of a spin-1/2 particle can be defined by the form
factors appearing in the decomposition of the matrix element of the
electromagnetic current~\cite{rev}:
\begin{eqnarray}
<f(p')|J_\mu|f(p)>=\bar{u}(p^\prime) \Gamma_{\mu}^{\gamma ff}(q)u(p) \;,
\label{eq:matrix}
\end{eqnarray}
where
\begin{eqnarray}
\Gamma^{\gamma f f}_\mu(q)=i e \biggl \{ \gamma_\mu \biggl [
V_f(q^2)-A_f(q^2) \gamma_5 \biggr ] + q^\nu \sigma_{\mu\nu}
\biggl [i \frac{\mu_f(q^2)}{e} - \frac{d_f(q^2)}{e} \biggl ] \biggr \} \;,
\end{eqnarray}
and $q=p^\prime-p$. This formula arises after making use of Lorentz
invariance, the Gordon identities and the fact that the external
photons and fermions are on-shell.  The operator which violates
CP-symmetry, ${\cal L}_{EDM}=-\frac{i}{2} d_f \bar{\psi}
\sigma_{\mu\nu}\gamma_5 \psi F^{\mu\nu}$ is non-renormalizable and of
dimension five. It reduces to the effective dipole interaction ${\cal
L}_{EDM}=d_f {\vec \sigma} \cdot {\vec E}$ in the non-relativistic
limit.

Experimental searches for electron and neutron EDMs currently provide
some of the most severe constraints on new models of particle physics:
\begin{eqnarray}
& &|d_e| \le 4.3\times 10^{-27} \,e{\rm  cm~\cite{electronedm}}, \\
& &|d_n| \le 6.3\times 10^{-26} \,e{\rm  cm~\cite{neutronedm}} \;.
\end{eqnarray}
All of the contributions to the EDMs or MDMs must be ultraviolate
finite because they are non-renormalizable interactions. In addition
the interactions flip the chirality of the external fermions and thus
break $SU(2)_L$ invariance. The chirality flip then comes from the
fermion masses which in turn come from the spontaneous breakdown of
electroweak gauge symmetry. By itself this is able to generate MDMs
but not EDMs for which CP-violation is needed.  In the SM the required
source of CP-violation resides in the complexity of the Yukawa
couplings which is parameterized by the CKM-phase\footnote{The main
contribution to the EDMs comes from the QCD $\theta$-angle. Here we
will assume (for alternatives see the discussion in ref.\cite{theta})
a Peccei-Quinn symmetry which is able to set this parameter to zero
(albeit at the price of an axion).}.  However the CKM-phase has only a
tiny contribution of $10^{-30}\,e$cm to the neutron
MDM~\cite{rev,kriplovich}. The CKM-phase can also penetrate the lepton
sector and generate EDMs for the leptons at higher loops but it has
been shown that these contributions vanish to three
loops~\cite{maxim}.  Consequently EDMs are a sensitive test of CP
violation beyond the SM.

If neutrinos are massive (as they indeed appear to be) then
CP-symmetry can also be violated in the leptonic sector and one
expects neutrino EDMs as well. These EDMs are induced by either a
CKM-like phase for Dirac neutrinos or by the three phases for Majorana
neutrinos in the leptonic mixing matrix.  Since one needs a chirality
flip for the EM vertex in order to generate EDMs, Majorana neutrinos
cannot have diagonal EDMs or MDMs. They can only have transition
electric or magnetic dipole moments~\cite{valle}, \ie a photon vertex
associated with two different neutrino flavours.  The experimental
bounds on the neutrino dipole moments are divided into two categories: 
The ``Earth bound'' constraints:
\begin{eqnarray}
& & |\mu_\nu| \le 1.5 \times 10^{-10}~\mu_B {\rm \cite{vogel}}, \\
& & |d_{\nu_\tau}| \le 5.2 \times 10^{-17}\,e{\rm cm~\cite{masso}} \;,
\end{eqnarray}
and the cosmological ones:
\begin{eqnarray}
& & |\mu_\nu| \le 3 \times 10^{-12}~\mu_B ~{\rm ~\cite{raffelt}},\\
& & |d_{\nu}| \le 2.5\times 10^{-22}~e{\rm cm~\cite{morgan}} \;.
\end{eqnarray}
These bounds can be used either for Dirac (diagonal EDMs or MDMs)
or for Majorana (transition EDMs or MDMs) neutrinos. 

In the SM the EDMs for the leptons due to a possible CP-violation in
the leptonic sector are too small to be significant. There is a
tendency for the various contributions to cancel or to be proportional
to $V_l V_l^*$.  The MDMs of the Dirac neutrinos in the SM with a
right handed singlet were calculated almost twenty years
ago~\cite{SMMDM}, and it was found that only a tiny loop induced
magnetic moment $\mu_\nu \simeq 3\times 10^{-19} \mu_B (m_\nu/1 {\rm
eV})$ arises.  Thus we conclude that in the SM the corrections to the
electron, neutron and neutrino EDMs and MDMs are very small and are
consistent with the data.

In the MSSM, apart from the CKM-phase or possible CP-violation in the
leptonic sector, there are additional sources of
CP-violation~\cite{masiero}. The soft breaking masses and couplings
can in general be complex, and their phases generate electron and
neutron EDMs even at the one-loop level in diagrams involving internal
squark/sleptons, charginos or gluinos~\cite{edmpapers}. The current
experimental bounds on the EDMs constrain those phases to be less than
$\sim 10^{-2}$, unless the phases are flavour off-diagonal, there are
cancellations between various contributions or the superpartner masses
are of order of 1 TeV~\cite{me}.  In the case of the neutrino MDM only
small corrections have been found in the MSSM with conserved R-parity
symmetry~\cite{MSSMMDM} and the result turns out to be similar to the
case of the SM ($\sim 10^{-19}~\mu_B$).

Models that violate R-parity can in principle induce additional
contributions to all these parameters \cite{review}. In this paper we
examine the effect of breaking R-parity on CP violating parameters.
We extend previous results by presenting a complete calculation of
electron, neutron and neutrino EDM/MDMs in the MSSM without R-parity
symmetry.
 
Before tackling the analysis in detail, we should make clear where our 
results differ from the previous estimates appearing in the literature. 
First, we show that any new contributions to the electron and neutron EDMs
are small and in fact appear only at two loops. In particular this means 
that constraints are only on products of 4 or more R-parity violating 
couplings. This result is in disagreement with the existing analysis
in the literature~\cite{frank}.  

Contributions to neutrino EDMs and MDMs can occur at one-loop,
however. Babu and Mohapatra~\cite{babu} were the first to consider the
MDM of the neutrino in the context of the MSSM with broken R-parity
symmetry.  They found contributions of order $\sim 10^{-11} \mu_B$
from the new loop graphs and thus a possible solution to the Solar
neutrino problem through the mechanism suggested in
ref.~\cite{voloshin}. We find results that are smaller than theirs by
a factor of 8. Barbieri et. al~\cite{barbieri} have also calculated
the neutrino MDMs and although we agree numerically with their result
to within an order of magnitude, we find that their formula for the
neutrino MDMs is unclear.  For example, it is not obvious from their
analysis that the diagonal neutrino MDM vanishes.  Finally, very
recently the neutrino MDMs were calculated in ref.\cite{pas} in a
notation (mass insertion) following closely that of
ref.\cite{barbieri}. We agree numerically to within an order of
magnitude with these previous estimates of the MDMs. Here we also
consider for the first time neutrino EDMs and determine the
corresponding bounds.

\section{Electron and neutron EDMs} 

Despite claims in the literature to the contrary~\cite{frank}, the
leading contributions to EDMs occur at two loops.  In this section we
show this using a combination of inspection and power counting
arguments.

First consider the extra contributions to the electron EDM from the
$\lambda LLE$ interactions. The EDM is found from the $q\rightarrow 0$
limit of $d_f$ in the matrix element of eq.(\ref{eq:matrix}). Hence
the relevant diagrams have one external left handed one external right
handed fermion and one photon.  Let us denote the number of chiral
superfields in the diagram by $n_L$ and $n_E$, and the number of
antichiral superfields by $n_{L^*}$ and $n_{E^*}$.  Adding $n_\lambda
$ of the LLE vertices to a diagram adds $2 n_\lambda $ to $n_L$ and
$n_\lambda $ to $n_E$. In addition we allow $D_L $ and $D_E$ of their
respective propagators. Each $L$ propagator removes one $L$ and one
$L^*$ and similarly for the $E$ propagators. Finally, we allow $D_m$
propagators with a mass insertion which changes the helicity on a
line. Each of these removes one $L^*$ and one $E^*$.  (We also allow
$D_{m^*}$ conjugate propagators).  Finally we note that any gauge
boson insertion do not change the number of $E,E^*,L$ or $L^*$.

Insisting that the final diagram has $n_L=1,~n_E=1,~n_{L^*}=0,~n_{E^*}=0$
yields four equations;
\ba 
n_L=1&=&2n_\lambda -D_L -D_{m^*}\nonumber \\
n_E=1&=&n_\lambda -D_E -D_{m^*}\nonumber \\
n_{L^*}=0&=&2n_{\lambda^*} -D_L -D_{m}\nonumber \\
n_{E^*}=0&=&n_{\lambda^*} -D_E -D_{m}
\ea
giving 
\ba
n_\lambda &=&n_{\lambda^*}\nonumber \\
D_m&=&D_{m^*}+1,
\ea
\ie we need at least one mass insertion to flip helicity.
Calling the number of non-gauge vertices $V = n_\lambda + 
n_{\lambda^*}$, the number of non-gauge 
propagators $I=D_L+D_E+D_m+D_{m^*}$ and the number of non-gauge 
external legs $E=2$, we 
can now use the standard power counting result
\ba
E &=& 3 V - 2 I \nonumber \\
L &=& I-V+1 
\ea
where $L$ is the number of loops. A little more algebra then gives 
\ba
L &=& n_\lambda \nonumber \\
I &=& 3 n_\lambda -1 .
\ea

\begin{figure}[htp]
\hspace*{+1.50in}
\epsfxsize=3.0in
\epsffile{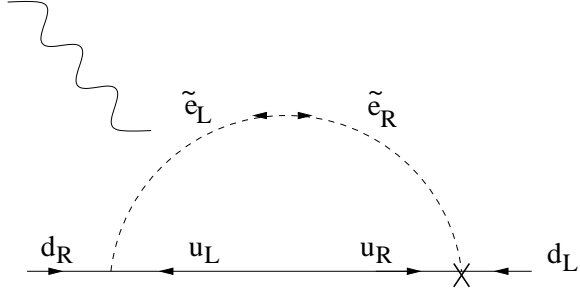}
\vspace{0.2in}
\caption{\it A one-loop diagram for neutron
EDMs showing explicitly the required helicity flips. 
This diagram does not contribute to the EDM in R-parity violating 
models of supersymmetry because of the absence of the crossed out vertex.} 
\end{figure}

Now consider one loop diagrams. The above tells us that they must have 
$n_\lambda=n_{\lambda^*}=1$ and that they must include at least one mass 
insertion. 
Inspection now shows that there are 
no {\em irreducible} diagrams of this kind that can give a contribution.
Indeed let us consider the contribution from the
apparently offending diagram shown in fig. 1. 
The relevant interaction terms come from
\ba
\label{leptonviol}
\delta L_ {L}  &=& \frac{1}{2}\lambda _{ijk} 
\left\{
\tilde{e}_{L_j} \overline{e}_k P_L \nu_i +
\tilde{\nu}_{L_i} \overline{e}_k P_L e_j + 
\tilde{e}^*_{R_k} \overline{\nu}^c_i P_L e_j  
- (i\leftrightarrow j) \right\} + \mbox{h.c.} \nn\\
  &+& \lambda' _{ijk} 
\left\{
\tilde{d}_{L_j} \overline{d}_k P_L \nu_i +
\tilde{\nu}_{L_i} \overline{d}_k P_L d_j + 
\tilde{d}^*_{R_k} \overline{\nu}^c_i P_L d_j  \right.
\nn \\ && \hspace{3cm}\left. - 
\tilde{u}_{L_j} \overline{d}_k P_L e_i -
\tilde{e}_{L_i} \overline{d}_k P_L u_j - 
\tilde{d}^*_{R_k} \overline{e}_i P_L u_j  
\right\} + \mbox{h.c.} \nn\\
\ea
where $P_L$ are projection operators.
In general, for Lagrangians of the form 
\be 
\phi^* \left( a \overline{\psi}_2 P_R \psi _1 + 
b \overline{\psi}_2 P_L \psi_1 \right)
\ee
where $\psi $ and $\phi$ are generic fermions and scalars, 
the one loop contributions to the fermion EDMs are 
proportional to Im($a^* b$) (see the third reference in \cite{edmpapers}). 
Thus the {\em same} scalar 
has to couple to both left and right helicities of a given fermion. 
For example, in the MSSM the one loop diagram with an internal 
chargino gives a contribution
to the down EDM because it contains both $\tilde{h}_1$ which couples to 
$\tilde{u}d_R$ {\em and} $\tilde{h}_2$ which couples to $\tilde{u}d_L$
(note the helicity flip required on the up-squark). There is 
an additional contribution that instead of $\tilde{h}_2$ involves 
the wino which also 
couples to $\tilde{u}d_L$. On the other hand the gluino 
gives a one loop contribution because it is a chargeless 
particle that can couple to both helicities thanks to its 
large Majorana mass. Indeed one can check that these 
three EDM contributions vanish if $\mu H_1 H_2 =0 $, $g_2=0$ and 
$m_{\tilde{g}}=0$.

Considering the R-parity violating one-loop diagrams with an internal
selectron, it is clear that, since there are no interactions that
involve $u_R$, there can be no contribution to the $u$ or $d$ EDMs.
(Note that there are extensions of the MSSM that {\em do} include 
such an interaction - however these also involve additional 
multiplets such as isosinglet down quarks coming from the {\bf 27} of 
$E_6$~\cite{grifols}.)
Likewise, $\lambda'$ only couples the electron to $u_L$ or $d_L$ (and
their conjugates) so that there is also no contribution to the
electron EDM from the $\lambda'$ vertex.  The diagrams with internal
(s)neutrinos can give EDMs only if (like the gluino) the neutrino has
a large ($\Delta L=2$) Majorana mass which of course it does not.
Finally we see that the only $P_R$ projection from the $\lambda$
vertex acts on the neutrino so that this vertex is also unable to
contribute to electron or quark EDMs.

In fact the first EDM contributions occur at two loops and hence
must have at least 4 $\lambda $ or $\lambda'$ vertices. 
Examples of the leading diagrams are shown in fig. 2, 
where the additional photon line may be attached to any internal 
(s)electron or (s)quark. 
The EDM can be found by extracting the leading 
linear term in $q$. For the example where the photon line is attached to 
the internal electron, we find 
\ba
\Gamma_\mu &=& e \sum_{ijlmn} m_l
\lambda_{1mn} \lambda_{jln} ^* \lambda_{iml} ^* \lambda_{ij1} ~
\overline{e}^\alpha e_\beta \times 
\nn\\&&
\int \int \frac{\dd ^4 p}{(2\pi)^4 }
\frac{\dd ^4 k}{(2\pi)^4 }\frac{1}{k^2-m_{\tilde{e}_L}^2}
\frac{1}{p^2-m_{\tilde{e}_R}^2} 
\nonumber \\
&& \times
q_\rho
\left\{ 
2 \gamma^\rho \gamma^\sigma \frac{p_\sigma (k+p)_\mu}
{k^2 ((k+p)^2-m_l^2 )^2 p^2}
+
4 \gamma_\mu \gamma^\sigma \frac{p_\sigma (k+p)_\rho}
{k^2 ((k+p)^2-m_l^2 )^3 }\right.
\nn\\&&
+
\left.
4 \gamma^\sigma \gamma^\nu \frac{p^\sigma k_\nu k^\rho 
(k+p)_\mu}
{k^4 ((k+p)^2-m_l^2 ) p^2 }
+
\gamma^\sigma \gamma_\mu \gamma^\rho \gamma^\nu
\frac{k^\sigma p_\nu}
{k^2 ((k+p)^2-m_l^2 )p^2 }
\right\}_{\alpha\gamma }P_{L_{\gamma \beta}}
\ea
The $F_3$ term comes by, for example, writing $\gamma^\mu\gamma^\nu 
= -i\sigma^{\mu\nu} + g^{\mu\nu}$.

\begin{figure}[tp]
\hspace*{+1.3in}
\epsfxsize=3.0in
\epsffile{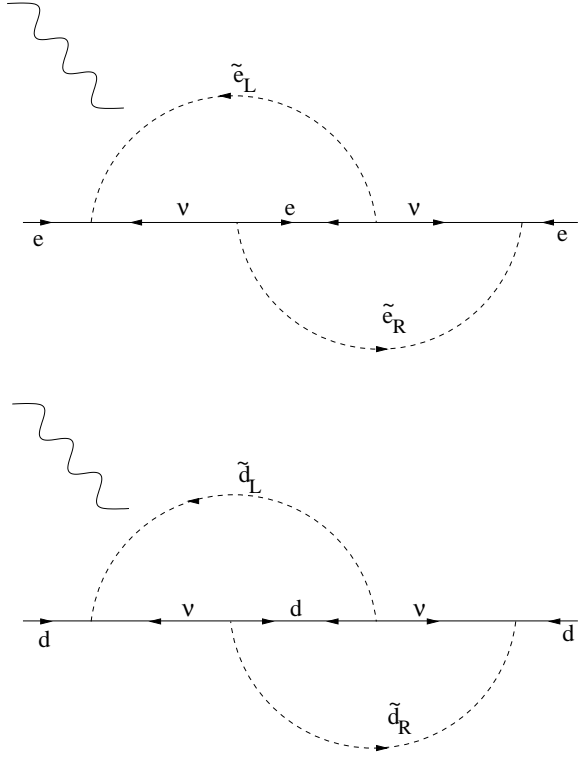}
\vspace{0.2in}
\caption{\it The leading 2 loop contributions to electron and neutron EDMs
in R-parity require at least two loops and 4 R-parity violating couplings.} 
\end{figure}

The evaluation of this integral by numerical methods is 
particularly difficult due to the presence of a kinematical 
singularity~\cite{broadhurst,avdeev}. Instead we note that the full 
calculation is similar to that of the anomalous magnetic moment of the 
muon presented in ref.\cite{avdeev}. For the present paper it 
is therefore sufficient to estimate the resulting EDM as
\be 
d_e = \sum_{ijlmn} \mbox{Im}\left( \frac{e }{(4\pi)^4  m_{\tilde{e}^2}} 
(a+b \log z_l)
m_l \lambda_{1mn} \lambda_{jln} ^* \lambda_{iml} ^* \lambda_{ij1}\right)
\ee
where
\be 
z_l= m_l^2 / m_{\tilde{e}^2} ,
\ee
and $a$ and $b$ are constants of ${\cal O}(1-10)$.
Putting in numbers we find that 
\be 
d_e \approx 
1.4 \times 10^{-22}e\mbox{\rm{cm}}
\left(\frac{100\gev}{m_{\tilde{e}}}\right)^2 
\times 
\mbox{Im} \left( 
 \sum_{ijlmn} (a+b \log(z_l))\frac{m_l}{m_\tau}
\lambda_{1mn} \lambda_{jln} ^* \lambda_{iml} ^* \lambda_{ij1}\right).
\ee
Comparing this number with the experimental constraint of 
$d_e < 10^{-28}$ \cite{electronedm} we find a bound 
\be
\mbox{Im} \left( 
 \sum_{ijlmn}\frac{m_l}{m_\tau}
\lambda_{1mn} \lambda_{jln} ^* \lambda_{iml} ^* \lambda_{ij1}\right) 
\leqsim 10^{-6},
\ee
where we conservatively take $a,b = 1$. 
The equivalent diagram for the neutrons yields 
\be 
d_n \approx 
1.4 \times 10^{-20}e\mbox{\rm{cm}}
\left(\frac{100\gev}{m_{\tilde{e}}}\right)^2 
\times 
\mbox{Im} \left( 
 \sum_{ijlmn} \frac{m_l}{m_t}
\lambda_{1mn} \lambda_{jln} ^* \lambda_{iml} ^* \lambda_{ij1}\right),
\ee
and comparing with experiment gives
\be
\mbox{Im} \left( 
 \sum_{ijlmn}\frac{m_l}{m_t}
\lambda_{1mn} \lambda_{jln} ^* \lambda_{iml} ^* \lambda_{ij1}\right) 
\leqsim 3\times 10^{-6}.
\ee
Given the strong bounds already existing on products of
couplings~\cite{bounds}, it is clear that these constraints from the
neutron and electron EDMs are far less important than
previously estimated~\cite{frank}.  In particular, since they
involve products of 4 couplings it is also clear that they
should easily be satisfied within any particular model.

\section{Neutrino MDMs and EDMs} 

\begin{figure}[htbp]
\begin{center}
\begin{tabular}{lr}
\begin{picture}(150,120)(0,0)
\ArrowLine(10,20)(30,20)
\Text(10,10)[]{\mbox{$\nu_i$}}
\Vertex(30,20){2}
\DashArrowLine(30,20)(90,20){7}
\Text(60,10)[]{\mbox{$\tilde{e}^-_{k_a}$}}
\Vertex(90,20){2}
\ArrowLine(110,20)(90,20)
\Text(110,10)[]{\mbox{${\nu^c_j}$}}
\ArrowLine(60,80)(30,20)
\Text(37,50)[r]{\mbox{$e_l^-$}}
\ArrowLine(90,20)(60,80)
\Text(85,50)[l]{\mbox{$e_l^-$}}
\Photon(60,80)(60,110){3}{3}
\Text(50,110)[c]{\mbox{$\gamma$}}
\end{picture}
&
\begin{picture}(150,120)(0,0)
\ArrowLine(10,20)(30,20)
\Text(10,10)[]{\mbox{$\nu_i$}}
\Vertex(30,20){2}
\ArrowLine(30,20)(90,20)
\Text(60,10)[]{\mbox{$e_l^-$}}
\Vertex(90,20){2}
\ArrowLine(110,20)(90,20)
\Text(110,10)[]{\mbox{${\nu^c_j}$}}
\DashArrowLine(60,80)(30,20){7}
\Text(37,50)[r]{\mbox{$\tilde{e}_{k_a}^-$}}
\DashArrowLine(90,20)(60,80){7}
\Text(85,50)[l]{\mbox{$\tilde{e}_{k_a}^-$}}
\Photon(60,80)(60,110){3}{3}
\Text(50,110)[c]{\mbox{$\gamma$}}
\end{picture}
\\
\begin{picture}(150,120)(0,0)
\ArrowLine(10,20)(30,20)
\Text(10,10)[]{\mbox{$\nu_i^c$}}
\Vertex(30,20){2}
\DashArrowLine(30,20)(90,20){7}
\Text(60,10)[]{\mbox{$\tilde{e}^+_{k_a}$}}
\Vertex(90,20){2}
\ArrowLine(110,20)(90,20)
\Text(110,10)[]{\mbox{${\nu_j}$}}
\ArrowLine(60,80)(30,20)
\Text(37,50)[r]{\mbox{$e_l^+$}}
\ArrowLine(90,20)(60,80)
\Text(85,50)[l]{\mbox{$e_l^+$}}
\Photon(60,80)(60,110){3}{3}
\Text(50,110)[c]{\mbox{$\gamma$}}
\end{picture}
&
\begin{picture}(150,120)(0,0)
\ArrowLine(10,20)(30,20)
\Text(10,10)[]{\mbox{$\nu_i^c$}}
\Vertex(30,20){2}
\ArrowLine(30,20)(90,20)
\Text(60,10)[]{\mbox{$e_l^+$}}
\Vertex(90,20){2}
\ArrowLine(110,20)(90,20)
\Text(110,10)[]{\mbox{${\nu_j}$}}
\DashArrowLine(60,80)(30,20){7}
\Text(37,50)[r]{\mbox{$\tilde{e}_{k_a}^+$}}
\DashArrowLine(90,20)(60,80){7}
\Text(85,50)[l]{\mbox{$\tilde{e}_{k_a}^+$}}
\Photon(60,80)(60,110){3}{3}
\Text(50,110)[c]{\mbox{$\gamma$}}
\end{picture}
%
\end{tabular}
\caption{\it Diagrams contributing to neutrino MDM and  EDM from the
R-parity violating coupling $\lambda_{ijk}L_iL_j\bar{E}_k$. The equivalent 
diagrams with the $\lambda^\prime_{ijk}L_iQ_j\bar{D}_k$ vertex are obtained 
by replacing $e$ with $d$.} 
\label{fig:edm_nu}
\end{center}
\end{figure}
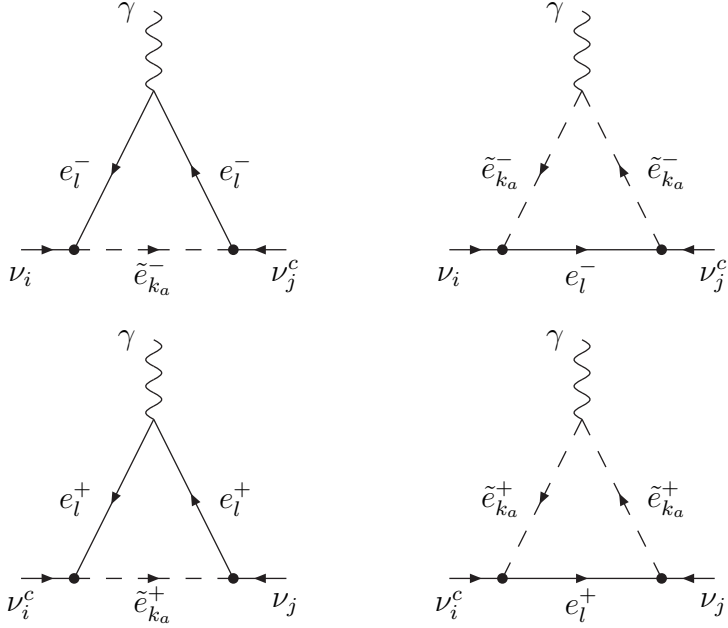

The exception in the discussion of the previous section was the
neutrino which {\em can} get E(M)DM contributions even at
one-loop.  The corresponding diagrams contributing to the neutrino MDM
and EDM are shown in fig.~\ref{fig:edm_nu}.

The neutrino MDMs and EDMs in the MSSM without R-parity can
be written as:
\begin{eqnarray}
\mu_{ij}^\nu \ &=& \ \frac{e}{32\pi^2} \Re e  \biggl \{
\sum_{a=1}^2 U^{\tilde{e}_k}_{1a} U^{*{\tilde{e}_k}}_{2a} 
\sum_{l,k=1}^3 \biggl (
 \lambda_{ikl}\lambda_{jlk}-\lambda_{jkl}\lambda_{ilk} \biggr )
m_{e_l} f(m_{e_l}^2,m_{\tilde{e}_{k_a}}^2) 
\nonumber \\
&+&
 \sum_{a=1}^2 U^{\tilde{d}_k}_{1a} U^{*{\tilde{d}_k}}_{2a} 
\sum_{l,k=1}^3 \biggl (
 \lambda^\prime_{ikl}\lambda^\prime_{jlk}-
\lambda^\prime_{jkl}\lambda^\prime_{ilk} \biggr )
m_{d_l} f(m_{d_l}^2,m_{\tilde{d}_{k_a}}^2) \biggr \} 
\label{nmdm} \\
d_{ij}^\nu \ &=& \ - \frac{e}{32\pi^2} \Im m \biggl \{ 
\sum_{a=1}^2 U^{\tilde{e}_k}_{1a}U^{*{\tilde{e}_k}}_{2a} 
\sum_{l,k=1}^3 \biggl (
 \lambda_{ikl}\lambda_{jlk}-\lambda_{jkl}\lambda_{ilk} \biggr )
m_{e_l} f(m_{e_l}^2,m_{\tilde{e}_{k_a}}^2) 
\nonumber \\
&+&
 \sum_{a=1}^2 U^{\tilde{d}_k}_{1a}U^{*{\tilde{d}_k}}_{2a} 
\sum_{l,k=1}^3 \biggl (
 \lambda^\prime_{ikl}\lambda^\prime_{jlk}-
\lambda^\prime_{jkl}\lambda^\prime_{ilk} \biggr )
m_{d_l} f(m_{d_l}^2,m_{\tilde{d}_{k_a}}^2) \biggr \}
\label{nedm}
\end{eqnarray}
where the function $f(x,y)$ is\footnote{We have made use of the approximation
$m_l\ll m_{\tilde{l}}$ and $m_q\ll m_{\tilde{q}}$.} 
\begin{equation}
f(x,y)= \frac{1}{y} \biggl ( 2+\ln \frac{y}{x} \biggr ) \;.
\end{equation}
The matrix $U^{\tilde{e}}$($U^{\tilde{d}}$) diagonalizes the slepton
(down squark) mass matrix and is given\footnote{Here we assume that
the soft SUSY CP-phases are small. Additional contributions to the
EDMs for the neutrinos are possible if they are large.} in terms of
the mixing angle $\theta_{\tilde{e}_i}$($\theta_{\tilde{d}_i}$)
\begin{eqnarray}
U^{\tilde{e}_i/\tilde{d}_i} = \biggl (
\begin{array}{cc}
\cos\theta_{\tilde{e}_i/\tilde{d}_i} & -\sin\theta_{\tilde{e}_i/\tilde{d}_i}
\\
\sin\theta_{\tilde{e}_i/\tilde{d}_i} & \cos\theta_{\tilde{e}_i/\tilde{d}_i}
\end{array} \biggr ) \;.
\end{eqnarray}
If the SUSY soft breaking masses of the slepton(squark) doublets,
$m_{\tilde{L}_i}$($m_{\tilde{Q}_i}$,$m_{\tilde{U}_i}$), 
are equal to those of the
right handed singlets, $m_{\tilde{e}_i}$($m_{\tilde{d}_i}$),
then to a very good approximation (the D-term contributions 
to the mixing angle are always small) we have: 
$\sin 2\theta_{\tilde{e}_i/\tilde{d}_i}\simeq 1$. Motivated also
by the fact that the bounds on R-parity violating couplings
are usually given using this simplification, we shall henceforth impose it. 
Thus, by expanding the sum over the
mass eigenstates of the sleptons(squarks) we obtain,
\begin{eqnarray}
\mu_{ij}^\nu \ &=& \ \frac{e}{64\pi^2} \Re e  \biggl \{
\sum_{l,k=1}^3 \biggl (
 \lambda_{ikl}\lambda_{jlk}-\lambda_{jkl}\lambda_{ilk} \biggr )
m_{e_l} \biggl [ f(m_{e_l}^2,m_{\tilde{e}_{k_1}}^2) -
f(m_{e_l}^2,m_{\tilde{e}_{k_2}}^2) \biggr ]
\nonumber \\
&+& 
\sum_{l,k=1}^3 \biggl (
 \lambda^\prime_{ikl}\lambda^\prime_{jlk}-
\lambda^\prime_{jkl}\lambda^\prime_{ilk} \biggr )
m_{d_l} \biggl [ f(m_{d_l}^2,m_{\tilde{d}_{k_1}}^2)-
 f(m_{d_l}^2,m_{\tilde{d}_{k_2}}^2) \biggr ] \biggr \} 
\label{nmdm2} \\
d_{ij}^\nu \ &=& \ - \frac{e}{64\pi^2} \Im m \biggl \{ 
\sum_{l,k=1}^3 \biggl (
 \lambda_{ikl}\lambda_{jlk}-\lambda_{jkl}\lambda_{ilk} \biggr )
m_{e_l} \biggl [f(m_{e_l}^2,m_{\tilde{e}_{k_1}}^2)-
 f(m_{e_l}^2,m_{\tilde{e}_{k_2}}^2) \biggr ] 
\nonumber \\
&+&
 \sum_{l,k=1}^3 \biggl (
 \lambda^\prime_{ikl}\lambda^\prime_{jlk}-
\lambda^\prime_{jkl}\lambda^\prime_{ilk} \biggr )
m_{d_l} \biggl [ f(m_{d_l}^2,m_{\tilde{d}_{k_1}}^2)-
f(m_{d_l}^2,m_{\tilde{d}_{k_2}}^2) \biggr ] \biggr \}
\label{nedm2}
\end{eqnarray}

Note that the diagrams of the first row in fig.~\ref{fig:edm_nu}
differ by a sign from those of the second row (due to the photon
vertex) and an interchange of the indices $i\leftrightarrow j$ (see
eqs.(\ref{nmdm},\ref{nedm})).

Some remarks are in order here:

\begin{itemize}

\item The diagonal elements of the MDMs and EDMs 
of the neutrinos are zero, \ie, $\mu^\nu_{ij}=d^\nu_{ij}=0$.
 This is of course a general
statement for the Majorana neutrinos. $i \ne j$ is assumed below.

\item For $k=l$ the contribution to eqs.(\ref{nmdm2},\ref{nedm2}) for
both the neutrino MDMs and EDMs are zero.

\item If one R-parity violating coupling dominates over the
others then again their contributions to the neutrino MDMs
and EDMs are zero. It is known~\cite{bounds} that even 
if we assume one coupling at a time at the GUT scale 
a number of lepton number violating couplings appear at the
electroweak scale since there is no symmetry (lepton symmetry)
to protect them. However,  we find that the effect on the
neutrino MDMs and EDMs is tiny \cite{rge}.

\item If the sleptons and  the squarks mass eigenstates are 
nearly degenerate then 
the MDM and EDM for the neutrinos
are much less than the experimental constraints.

\end{itemize}

If there are no other CP-violating sources (such as SUSY CP-phases)
apart from the CKM-phase then one might still 
expect some transmission of this
phase into the EDMs of the neutrinos.
Here we prove that there is no such effect. Without loss of
generality,
we  assume that the
CP-violating phase appears in the CKM through the down 
quark Yukawa couplings. Then after the redefinition of the 
fields~\cite{agashe,bounds} we obtain
\begin{equation}
\lambda_{ijk}^\prime = \tilde{\lambda}_{ilm}^\prime (V_{CKM})_{mk}
(V_{CKM}^\dagger)_{jl},
\label{redef}
\end{equation}
where $\tilde{\lambda}_{ilm}^\prime$ in the right hand side is a real
coupling, whence 
\begin{eqnarray}
\Im m \sum_{l,k=1}^3 
\biggl (
 \lambda^\prime_{ikl}\lambda^\prime_{jlk}-
\lambda^\prime_{jkl}\lambda^\prime_{ilk} \biggr )= \Im m
\biggl (
 \tilde{\lambda}^\prime_{ikl}\tilde{\lambda}^\prime_{jlk}-
\tilde{\lambda}^\prime_{jkl}\tilde{\lambda}^\prime_{ilk} \biggr ) = 0 \;.
\end{eqnarray}
Hence there are no neutrino EDMs coming from the CKM-phase contribution.
Following precisely the same arguments we can prove that
the neutrino EDMs are zero even if we make the assumption 
that there is  CP-violation in
the leptonic sector (from the three Majorana phases).

We now consider the contributions to neutrino MDMs. 
The importance of each term in eq.(\ref{nmdm2})
depends on which are the dominant R-parity violating couplings and 
also on the degeneracy of the slepton and squark mass 
eigenstates. Here we shall assess the maximum contribution
of the RPV couplings to the neutrino MDMs by
taking one of the two slepton/squark mass eigenstates
e.g $m_{\tilde{e}_2},m_{\tilde{d}_2}$ to be in
the decoupling region (the function $f(x,y)$ goes to zero for large 
$y$ ). Beacom and Vogel~\cite{vogel} have recently shown
that for Majorana neutrinos with two flavours the neutrino MDMs
are given by
\begin{equation}
\mu_{\nu_e}^2=|\mu_{12}|^2 \;,
\end{equation}
for either vacuum or MSW mixing, and the bound obtained from 
SuperKamiokande solar neutrino data~\cite{superK} is
\begin{equation}
|\mu_{\nu_e}| \le 1.5\times 10^{-10} \mu_B \;\; (90\% CL) \;.
\end{equation}
By using this bound and assuming that the 
sleptons and squarks of each generation are almost degenerate \ie, 
$m_{\tilde{e}}=m_{\tilde{\mu}}=m_{\tilde{\tau}}$ and
$m_{\tilde{d}}=m_{\tilde{s}}=m_{\tilde{b}}$ we find
\begin{eqnarray}
& &\lambda_{121}\lambda_{212}\biggl ( m_e f(m_e^2,m_{\tilde{e}}^2)
-m_\mu  f(m_\mu^2,m_{\tilde{e}}^2) \biggr ) \nonumber \\ + & &
\lambda_{131}\lambda_{213}\biggl ( m_e f(m_e^2,m_{\tilde{e}}^2)
-m_\tau  f(m_\tau^2,m_{\tilde{e}}^2) \biggr ) \nonumber \\ + & &
\lambda_{123}\lambda_{232}
\biggl ( m_\tau f(m_\tau^2,m_{\tilde{e}}^2)-
m_\mu  f(m_\mu^2,m_{\tilde{e}}^2) \biggr ) \nonumber \\ + & &
 \biggl (\lambda^\prime_{121}\lambda^\prime_{212} -
\lambda^\prime_{221}\lambda^\prime_{112} \biggr )
\biggl ( m_d f(m_d^2,m_{\tilde{d}}^2)
-m_s  f(m_s^2,m_{\tilde{d}}^2) \biggr ) \nonumber \\ + & &
 \biggl (\lambda^\prime_{131}\lambda^\prime_{213} -
\lambda^\prime_{231}\lambda^\prime_{113} \biggr )
\biggl ( m_d f(m_d^2,m_{\tilde{d}}^2)
-m_b  f(m_b^2,m_{\tilde{d}}^2) \biggr ) \nonumber \\ + & &
 \biggl (\lambda^\prime_{132}\lambda^\prime_{223} -
\lambda^\prime_{123}\lambda^\prime_{232} \biggr )
\biggl ( m_s f(m_s^2,m_{\tilde{d}}^2)
-m_b  f(m_b^2,m_{\tilde{d}}^2) \biggr ) \le 10^{-4} \;.
\label{mag}
\end{eqnarray}
For $m_{\tilde{e}}=m_{\tilde{d}}=100$~GeV and one
dominant pair of R-parity violating couplings at a
time we obtain the following bounds:
\begin{eqnarray}
\Re e (\lambda_{121}\lambda_{212}) & < & 0.58 \label{b1}\\
\Re e (\lambda_{131}\lambda_{213}) & < & 0.059 \\
\Re e (\lambda_{123}\lambda_{232}) & < & 0.063\\
\Re e (\lambda^\prime_{121}\lambda^\prime_{212}) \;,\;
\Re e (\lambda^\prime_{221}\lambda^\prime_{112}) & < & 0.60 \\
\Re e (\lambda^\prime_{131}\lambda^\prime_{213}) \;,\;
\Re e (\lambda^\prime_{231}\lambda^\prime_{113}) \;,\;
\Re e (\lambda^\prime_{132}\lambda^\prime_{223}) \;,\;
\Re e (\lambda^\prime_{123}\lambda^\prime_{232}) & < & 0.030 \;.
\label{bounds}
\end{eqnarray}
If we now compare these bounds with those shown in ref.\cite{bounds},
we find that they are all far more relaxed
than the constraints obtained from  other processes
(even in some cases more relaxed than the individual bounds on 
the corresponding RPV couplings). 
Thus we conclude that the contribution from the R-parity
violating couplings to the neutrino MDMs is rather small.

It is possible in general to start with complex RPV
couplings. Assuming no neutrino mixing here the induced neutrino
EDMs\footnote{There are no neutrino MDMs in this case.} are given
by~\cite{raffelt}
\begin{equation}
d^\nu =\frac{1}{2}
 \sum_{i,j=1}^3 |d_{ij}|^2 = |d_{12}|^2+|d_{13}|^2+|d_{23}|^2 \;.
\end{equation}
By considering the cosmological bound\footnote{The accelerator
bound of \cite{masso} is almost five orders of magnitude
more relaxed than the cosmological one and does not constrain
the RPV couplings at all.} of~\cite{morgan}
$d^\nu=2.5\times 10^{-22}\,e{\rm cm}$ and assuming one dominant 
pair of RPV couplings at a time we obtain 
($m_{\tilde{e}}=m_{\tilde{d}}=100$~GeV):
\begin{eqnarray}
\Im m (\lambda_{i21}\lambda_{j12}) & < & 0.05 \\
\Im m (\lambda_{i31}\lambda_{j13}) & < & 0.004 \\
\Im m (\lambda_{i23}\lambda_{j32}) & < & 0.005\\
\Im m (\lambda^\prime_{i21}\lambda^\prime_{j12}) 
 & < & 0.06 \\
\Im m (\lambda^\prime_{i31}\lambda^\prime_{j13})  \;,\;
\Im m (\lambda^\prime_{i32}\lambda^\prime_{j23}) 
 & < & 0.0024 \;.
\label{edmnds}
\end{eqnarray}
These new EDM results conclude this chapter and the discussion 
on neutrino MDMs and EDMs.

\section{Conclusions}
We have shown that in the MSSM without R-parity symmetry it is
impossible to generate additional electron and neutron EDMs at 1-loop
from the R-parity violating Yukawa couplings. EDMs for the electron
and neutron first arise at the 2-loop level and we estimated the
contribution of the new two-loop graphs. We find that the resulting
constraints are on products of at least 4 R-parity violating Yukawa
couplings; 
\ba 
\mbox{Im} \left( \sum_{ijlmn}\frac{m_l}{m_\tau}
\lambda_{1mn} \lambda_{jln} ^* \lambda_{iml} ^* \lambda_{ij1}\right)
&\leqsim & 10^{-6},\nn\\ \mbox{Im} \left( \sum_{ijlmn}\frac{m_l}{m_t}
\lambda_{1mn} \lambda_{jln} ^* \lambda_{iml} ^* \lambda_{ij1}\right)
&\leqsim & 3\times 10^{-6}.  
\ea 
Conversely we find that (Majorana) neutrino electric and magnetic
transition moments are non-zero even at the 1-loop level. Constraints
on the R-parity violating couplings were derived using the current
bounds on neutrino dipole moments.

\section*{Acknowledgments}

A.D. is funded by Marie Curie Research Training Grant
ERB-FMBI-CT98-3438. 

\bigskip

\noindent {\bf Note added in proof:} Whilst in the final stages of 
preparation ref.\cite{godbole} appeared which draws the same conclusion 
concerning the electron and neutron EDMs.


\begin{thebibliography}{99}

\bibitem{rev}
W.~Bernreuther and M.~Suzuki,
Rev.\ Mod.\ Phys.\ {\bf 63} (1991) 313.

\bibitem{electronedm}
K.~Abdullah et al., \prl {65}{2340}{1990};
E.~Commins et al., Phys.~Rev.~{\bf A~50} (1994) 2960;
B.E.~Sauer, J.~Wang and E.A.~Hinds, \prl{74}{1554}{1995};
{\it J. Chem. Phys} {\bf 105} (1996) 7412.

\bibitem{neutronedm}
P.G.~Harris et al., \prl{82}{904}{1999}.

\bibitem{theta}
A.~Dedes and M.~Pospelov,
{\tt hep-ph/9912293}.

\bibitem{kriplovich}
I.B.~Khriplovich and S.K.~Lamoreaux, {\it ``CP Violation 
Without Strangness''}, Springer 1997.

\bibitem{maxim}
M. Pospelov, private communication.

\bibitem{valle}
L.~Wolfenstein, \plb{107}{77}{1981};
J.~Schechter and J.W.F.~Valle, \prd{24}{1883}{1981};
B.~Kayser, \prd{26}{1662}{1982};
R.~Shrock, \npb{206}{359}{1982};
J.F.~Nieves, \prd{26}{3152}{1982}.

\bibitem{vogel}
J.F.~Beacom and P.~Vogel,
{\tt hep-ph/9907383}.

\bibitem{masso}
R.~Escribano and E.~Masso,
Phys.\ Lett.\ {\bf B395} (1997) 369
{\tt hep-ph/9609423}.

\bibitem{raffelt}
G.G.~Raffelt,
Phys.\ Rev.\ Lett.\ {\bf 64} (1990) 2856.

\bibitem{morgan}
J.A.~Morgan and D.B.~Farrant,
Phys.\ Lett.\ {\bf B128} (1983) 431.

\bibitem{SMMDM}
B.W.~Lee and R.E.~Shrock,
Phys.\ Rev.\ {\bf D16} (1977) 1444;
W.J.~Marciano and A.I.~Sanda,
Phys.\ Lett.\ {\bf 67B} (1977) 303.

\bibitem{masiero}
For a review, see, \eg:\\
A.~Masiero and L.~Silvestrini, lecture given at the `International
School on Subnuclear Physics, 35th Course', Erice, Italy, 1997,
{\tt hep-ph/9711401}.

\bibitem{edmpapers}
Y.~Kizukuri and N.~Oshimo,
Phys.\ Rev.\ {\bf D46} (1992) 3025;
S.~Bertolini and F.~Vissani,
Phys.\ Lett.\ {\bf B324} (1994) 164
{\tt hep-ph/9311293};
S.A.~Abel, W.N.~Cottingham and I.B.~Whittingham,
Phys.\ Lett.\  {\bf B370} (1996) 106
[hep-ph/9511326];
A.~Romanino and A.~Strumia,
Nucl.\ Phys.\ {\bf B490} (1997) 3;
T.~Ibrahim and P.~Nath, Phys. Rev. {\bf D57} (1998) 478;
T.~Falk and K.A.~Olive, Phys. Lett. {\bf B439} (1998)  71;
T.~Falk, A.~Ferstl and K.A.~Olive, Phys. Rev. {\bf D59} (1999) 055009;
T.~Ibrahim and P.~Nath, Phys. Rev. {\bf D58} (1998)  111301;
M.~Brhlik, G.J.~Good and G.L.~Kane, Phys. Rev. {\bf D59} (1999) 115004;
A.~Bartl, T.~Gajdosik, W.~Porod, P.~Stockinger and H.~Stremnitzer, preprint
UWThPh--1998--63, HEPHY--PUB 705, March 1999, {\tt hep-ph/9903402};
T.~Falk, K.A.~Olive, M.~Pospelov and R.~Roiban, {\tt hep-ph/9904393}.
S.~Pokorski, J.~Rosiek and C.A.~Savoy, 
{\tt hep-ph/9906206}; Some of the two loop contributions
to the electron and neutron EDMs in the MSSM have been found in:
D.~Chang, W.~Keung and A.~Pilaftsis,
Phys.\ Rev.\ Lett.\ {\bf 82} (1999) 900
{\tt hep-ph/9811202}; A.~Pilaftsis,
{\tt hep-ph/9909485}; D.~Chang, W.~Chang and W.~Keung,
{\tt hep-ph/9910465}; A.~Pilaftsis,
{\tt hep-ph/9912253}.

\bibitem{me}
S.~Dimopoulos and G.~Giudice, \plb{357}{573}{1995};
S.A.~Abel and J.-M.~Fr\`ere \prd{55}{1623}{1997}; 
S.A.~Abel, \plb{410}{173}{1997};
S.~Khalil, T.~Kobayashi and A.~Masiero,
Phys.\ Rev.\  {\bf D60} (1999) 075003
[hep-ph/9903544];
G.~C.~Branco, F.~Cagarrinho and F.~Kruger,
Phys.\ Lett.\  {\bf B459} (1999) 224
[hep-ph/9904379];
D.A.~Demir,
Phys.\ Rev.\  {\bf D60} (1999) 095007
[hep-ph/9905571];
S.~Khalil and T.~Kobayashi,
Phys.\ Lett.\  {\bf B460} (1999) 341
[hep-ph/9906374];
M.~Brhlik, L.~Everett, G.~L.~Kane and J.~Lykken,
Phys.\ Rev.\ Lett.\  {\bf 83} (1999) 2124
[hep-ph/9905215];
M.~Brhlik, L.~Everett, G.~L.~Kane, S.~F.~King and O.~Lebedev,
hep-ph/9909480;
see 
Y.~Nir,
hep-ph/9911321.

\bibitem{MSSMMDM}
S.N.~Biswas, A.~Goyal and J.N.~Passi,
Phys.\ Rev.\ {\bf D28} (1983) 671;
T.~M.~Aliev,
Yad.\ Fiz.\ {\bf 44} (1986) 1043;
J.~Liu,
Phys.\ Rev.\ {\bf D35} (1987) 3447.
K.L.~Ng,
Z.\ Phys.\ {\bf C48} (1990) 289.

\bibitem{review}
For a review on R-parity violation see, {\it e.g.}
H. Dreiner, hep-ph/9707435; G. Bhattacharyya, Nucl. Phys. Proc. 
Suppl. 52A (1997) 83, hep-ph/9608415; {\it ibid} hep-ph/9709395.

\bibitem{frank}
M.~Frank and H.~Hamidian,
J.\ Phys.\ {\bf G24} (1998) 2203,
hep-ph/9706510.

\bibitem{babu}
K.S.~Babu and R.N.~Mohapatra,
Phys.\ Rev.\ Lett.\ {\bf 64} (1990) 1705.

\bibitem{voloshin}
L.B.~Okun, M.B.~Voloshin and M.I.~Vysotsky,
Sov.\ J.\ Nucl.\ Phys.\ {\bf 44} (1986) 440;
R. Cisneros, Astrophys. Space Sci. {\bf 10}, 87 (1971).

\bibitem{barbieri}
R.~Barbieri, M.M.~Guzzo, A.~Masiero and D.~Tommasini,
Phys.\ Lett.\ {\bf B252} (1990) 251.

\bibitem{pas}
G.~Bhattacharyya, H.V.~Klapdor-Kleingrothaus and H.~Pas,
Phys.\ Lett.\ {\bf B463} (1999) 77,
hep-ph/9907432.

\bibitem{grifols}
J.~A.~Grifols and J.~Sola,
Phys.\ Lett.\  {\bf B189} (1987) 63

\bibitem{broadhurst}
P.A.Baikov and D.J.Broadhurst, {\em New Computing techniques in 
Physics Research} {\bf IV} (1995) 167.
 
\bibitem{avdeev}
L.V.Avdeev, J.Fleischer, M.Yu.Kalmykov and M.N.Tentyukov, 
{\em Comput. Phys. Commun.} {\bf 107} (1997) 155.

\bibitem{bounds}
B.C.~Allanach, A.~Dedes and H.K.~Dreiner,
Phys.\ Rev.\ {\bf D60} (1999) 075014
hep-ph/9906209.

\bibitem{rge} For this we have made use of the computations described in
B.C. Allanach, A. Dedes, H.K. Dreiner, Phys. Rev. D60 (1999) 056002, 
hep-ph/9902251. 

\bibitem{agashe}
K.~Agashe and M.~Graesser,
Phys.\ Rev.\ {\bf D54} (1996) 4445
hep-ph/9510439.

\bibitem{superK}
SuperKamiokande Collaboration, Y. Fukuda et.al., \prl{81}{1998}{1562}.


\bibitem{godbole}
R.~M.~Godbole, S.~Pakvasa, S.~D.~Rindani and X.~Tata,
hep-ph/9912315.
\end{thebibliography}
\end{document}